\documentclass[aip,apl,twocolumn,superscriptaddress,reprint]{revtex4-1}
\usepackage{amssymb}
\usepackage{amsmath}
\usepackage{graphicx}
\usepackage{natbib}
\usepackage{epstopdf}
\usepackage{color}

\begin{document}

\title{Circuit Quantum Electrodynamics Architecture for Gate-Defined Quantum Dots in Silicon}
\author{X. Mi}
\affiliation{Department of Physics, Princeton University, Princeton, New Jersey 08544, USA}
\author{J. V. Cady}
\altaffiliation{Present Address: Department of Physics, University of California Santa Barbara, Santa Barbara, California 93106, USA}
\affiliation{Department of Physics, Princeton University, Princeton, New Jersey 08544, USA}
\author{D. M. Zajac}
\affiliation{Department of Physics, Princeton University, Princeton, New Jersey 08544, USA}
\author{J. Stehlik}
\altaffiliation{Present Address: IBM T.J. Watson Research Center, Yorktown Heights, New York 10598, USA}
\affiliation{Department of Physics, Princeton University, Princeton, New Jersey 08544, USA}
\author{L. F. Edge}
\affiliation{HRL Laboratories LLC, 3011 Malibu Canyon Road, Malibu, California 90265, USA}
\author{J. R. Petta}
\affiliation{Department of Physics, Princeton University, Princeton, New Jersey 08544, USA}

% 03.67.Lx Quantum computation
% 73.21.La Quantum dots
% 42.50.Pq Cavity quantum electrodynamics; micromasers
% 85.35.Gv Single electron devices

\begin{abstract}
We demonstrate a hybrid device architecture where the charge states in a double quantum dot (DQD) formed in a Si/SiGe heterostructure are read out using an on-chip superconducting microwave cavity. A quality factor $Q$ = 5,400 is achieved by selectively etching away regions of the quantum well and by reducing photon losses through low-pass filtering of the gate bias lines. Homodyne measurements of the cavity transmission reveal DQD charge stability diagrams and a charge-cavity coupling rate $g_\text{c} / 2\pi =$ 23 MHz. These measurements indicate that electrons trapped in a Si DQD can be effectively coupled to microwave photons, potentially enabling coherent electron-photon interactions in silicon.
\end{abstract}

\maketitle
Silicon is an emerging material system for spin-based quantum computing due to record long quantum coherence times. \cite{Lyon_NatMat_2012,Saeedi2013} Spin states of electrons in semiconductor quantum dots (QDs), long recognized to be highly promising candidates for the storage of quantum information, \cite{Loss_DiVincenzo_PRA} have limited coherence times in traditional host materials such as GaAs due to fluctuations of the nuclear spin bath. \cite{Petta_Science,Petta_RevMod} In silicon, owing to the zero nuclear spin carried by the naturally abundant isotope $^{28}$Si, hyperfine induced dephasing of electron spins is strongly reduced. \cite{Erik_QIP} In contrast with III/V semiconductor compounds, silicon has weak spin-orbit coupling and can be isotopically enriched to the level of 800 ppm $^{29}$Si for further enhancement of spin coherence times. \cite{Sturm_APL2012}

Quantum devices based on Si can potentially be scaled to larger system sizes using well-developed semiconductor fabrication processes. Recent advances include the demonstration of two-qubit logic gates \cite{Veldhorst2015} and the fabrication of a one-dimensional chain of nine QDs that was measured using three proximal charge detectors. \cite{Dave_Ninedot_PhysRevApp} Moreover, the long coherence times that have been reported in Si pave the way for long range coupling of spin states using superconducting cavities in the circuit quantum electrodynamics (cQED) architecture. \cite{Sillanpaa2007,Majer2007}

\begin{figure}[t]
	\centering
	\includegraphics[width=\columnwidth]{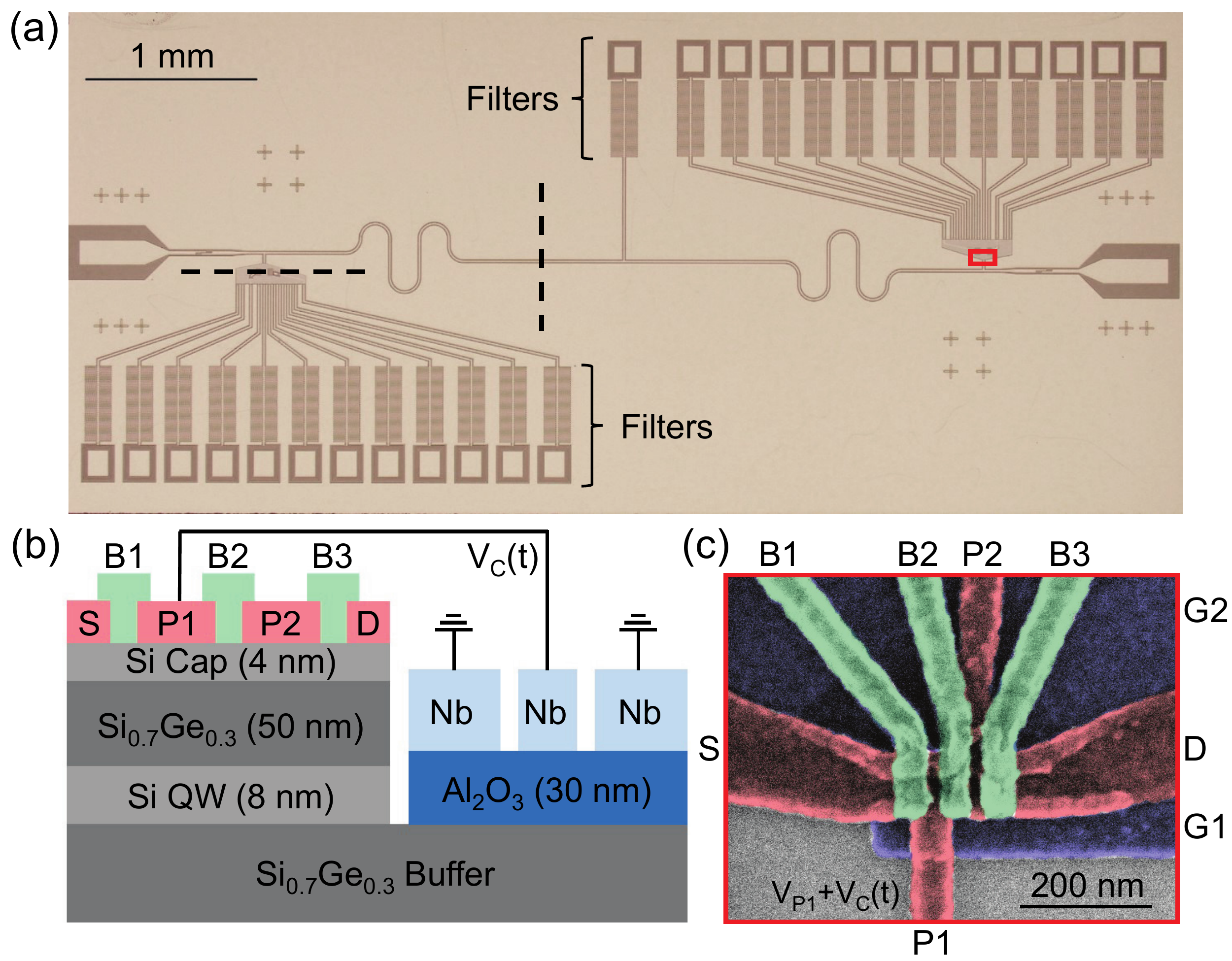}
	\caption{(a) Optical image of a silicon hybrid cQED device. A Si DQD is placed at each voltage anti-node of the cavity. $LC$ filters reduce leakage of cavity photons through dc biasing lines. (b) A cross-section (not to scale) taken along the horizontal dashed line in (a) shows the overlapping Al gates that define the DQD and the Si/SiGe heterostructure layers. To minimize internal losses the quantum well is removed in areas beneath the cavity center pin [cross-section through the vertical dashed line in (a)].  (c) False-color scanning electron microscope image of a DQD.}
	\label{fig:1}
\end{figure}

The field of cQED experimentally realizes on-chip interactions between a two-level system (the qubit) and photons confined within a superconducting microwave cavity. \cite{Strong_Coupling_CooperPair} Such cavities typically have frequencies between 1 and 10 GHz, which match the transition frequencies of many nanofabricated quantum devices and are therefore suitable mediators of non-local qubit interactions, providing a means for long range scaling of solid state qubits. In cQED systems with superconducting qubits, cavity photons are also widely used for dispersive state readout, as the significant electric dipole moments of these devices result in large phase shifts in the cavity response. \cite{DiCarlo,Vijay}  In semiconductor systems, hybrid cQED devices have been implemented using GaAs, \cite{Walraff_2012_PRL,Toida_GaAs_PRL} InAs, \cite{Petersson_Nature_2012} carbon nanotube, \cite{Viennot408} and graphene QDs. \cite{Graphene_cQED_2015} There are several proposals pertaining to the coupling of Si spin qubits to cavities, \cite{Taylor_PRL_2013,Guido_RX_PRB_2015} as well as the demonstration of a high kinetic inductance cavity, fabricated with the intention of coupling to Si quantum dots. \cite{Vandersypen_HighKinetic}

In this Letter, we present a hybrid cQED device architecture that couples a silicon DQD to a superconducting cavity. The device has three key components: a half-wavelength co-planar waveguide (CPW) cavity, two gate-defined DQDs, and low-pass $LC$ filters that serve to reduce microwave losses through the dc bias lines that are used to tune DQD. This paper is organized as follows. We describe the device layout and fabrication process, the $LC$ filter design considerations, and then demonstrate readout of DQD charge states using the cavity. 

The devices are fabricated on Si/SiGe heterostructures grown by chemical vapor deposition. \cite{Payette_APL,Mi_Hall_PRB} A 3 $\mu$m thick linearly graded Si$_{1-x}$Ge$_x$ relaxed buffer substrate is grown on top of a Si wafer (resistivity $>5000$ $\Omega$-cm). The buffer is chemically and mechanically polished before the growth of a 170 -- 375 nm thick Si$_{0.7}$Ge$_{0.3}$ layer, an 8 nm thick Si quantum well (QW), a 50 -- 60 nm thick Si$_{0.7}$Ge$_{0.3}$ spacer and a 2 -- 4 nm thick Si cap. Wafers grown under similar conditions have maximum mobilities $\mu =$ 650,000 cm$^2$/Vs and can support electron densities up to $n = 8 \times 10^{11}/$cm$^2$.

The cavity fabrication process is designed to achieve two major goals: protection of the Si QW from the reactive ions used to etch the Nb cavity and the reduction of internal photon losses introduced by two-level system (TLS) defects at the  heterostructure interfaces. The Si QW in the area under the cavity center pin is first removed through a 70 nm deep reactive ion etch to minimize the internal photon loss. A 30 nm thick Al$_2$O$_3$ film, which serves as an etch stop for cavity fabrication, is then grown over the entire substrate using atomic layer deposition. Next, a 50 nm thick Nb film is deposited via dc sputtering. The cavity and filter patterns, shown in Fig.~\ref{fig:1}(a), are defined with a second reactive ion etch step using a SF$_6$/Ar plasma. A hydrofluoric acid etch then removes the Al$_2$O$_3$ film in the area where the DQD is to be defined. The resulting cross-sections of the device are schematically illustrated in Fig.~\ref{fig:1}(b).

Accumulation-mode DQDs are defined using an overlapping gate architecture. \cite{Dave_Ninedot_PhysRevApp} A scanning electron microscope image of the DQD is shown in Fig.~\ref{fig:1}(c). For the measurements presented here, electrons are only accumulated in one of the DQDs and the other DQD does not contribute to the cavity response. The first Al layer, shaded in purple, consists of two large gates G1 and G2 that selectively screen the electrostatic potentials of the upper Al layers and form a quasi-one-dimensional transport channel. The second Al layer, shaded in pink, consists of two plunger gates P1 and P2 that are used to tune the chemical potentials of the DQD, as well as source (S) and drain (D) accumulation gates. P1 is connected to the cavity center pin and capacitively couples the DQD to the time-dependent voltage $V_\text{C} (t)$ of the cavity. A dc tap is placed at the voltage node of the cavity and used to dc bias gate P1. The third Al layer, shaded light green, consists of three tunnel barrier gates. Gate B2 tunes the interdot tunnel coupling ($t_\text{c}$), gate B1 tunes the dot 1-- source reservoir coupling, while gate B3 tunes the dot 2--drain reservoir coupling. The dots have average charging energies $E_\text{c}$ = 6.9 $\pm$ 0.7 meV, average orbital energies $E_\text{orb}$ = 3.0 $\pm$ 0.5 meV, and valley splittings in the range of 35--70 $\mu$eV.\cite{Dave_DQD_APL,Dave_Ninedot_PhysRevApp}

\begin{figure}[t]
	\centering
	\includegraphics[width=\columnwidth]{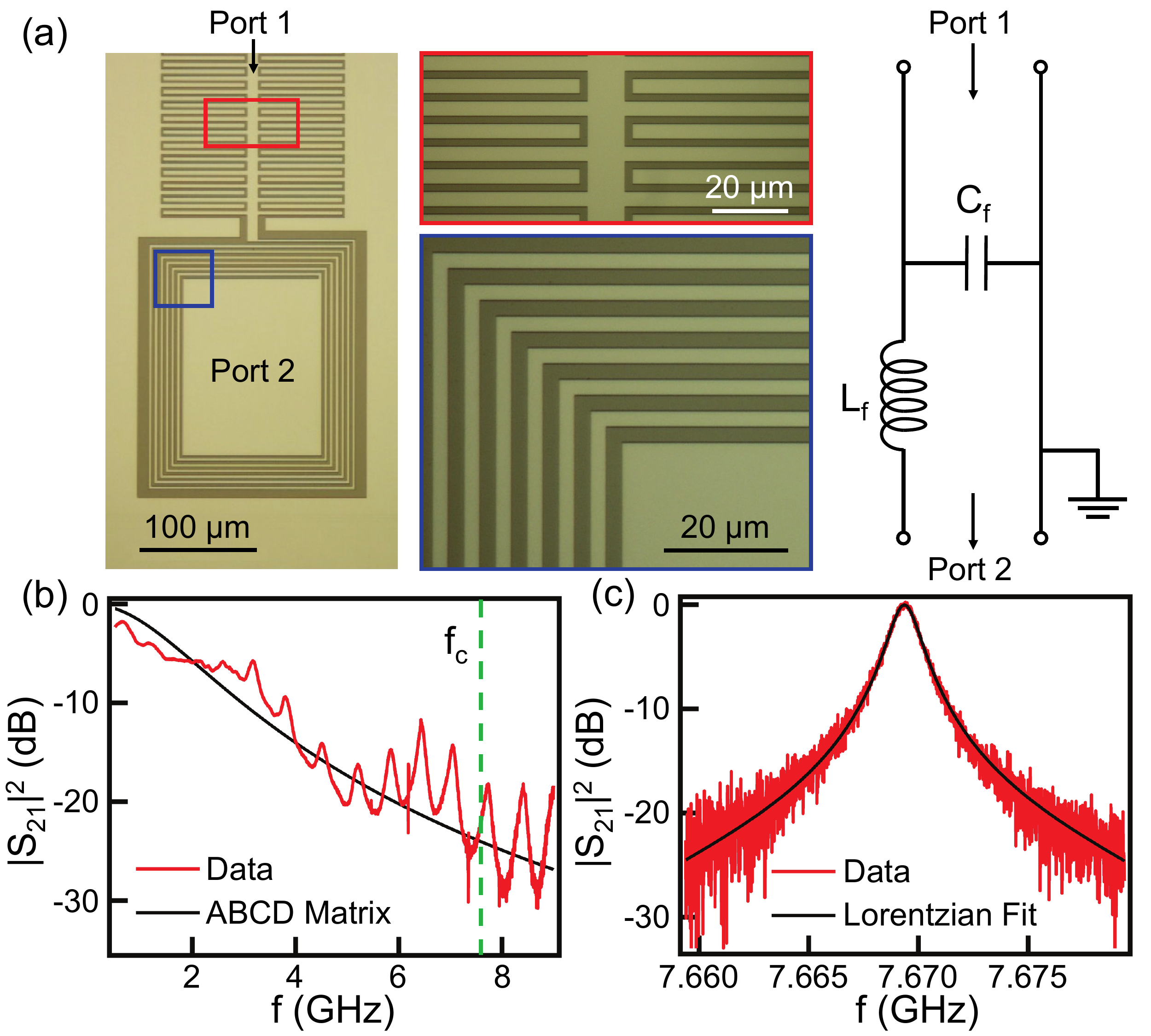}
	\caption{(a) Left: Optical image of a compact $LC$ filter, showing the spiral inductor and a portion of the capacitor. Middle: Zoomed-in view of the capacitor (red outline)/inductor (blue outline). Right: Circuit model for the $LC$ filter. (b) Measured $LC$ filter transmission $|S_{21}|^2$ (red) and ABCD matrix predictions (black). (c) Cavity transmission, $|S_{21}|^2$, measured with the DQD in Coulomb blockade. The black line is a fit to a Lorentzian with $Q = 5400$.}
	\label{fig:2}
\end{figure}

In comparison with superconducting qubit cQED devices, which generally support $Q$ $>$ 50,000, \cite{Quintana_APL} hybrid QD-cQED systems generally have $Q$ = 1,000 -- 3,000. \cite{Walraff_2012_PRL,Petersson_Nature_2012,Graphene_cQED_2015,Toida_GaAs_PRL} To coherently couple QD qubits to cavity photons, higher quality factors are needed. Measurements on previously reported device designs \cite{Petersson_Nature_2012} show significant microwave leakage through the dc bias lines leading to the DQD, which we attribute to the capacitive coupling between the cavity and each of the dc bias lines. The bias lines therefore become leakage pathways that inadvertently lower the $Q$.  To minimize microwave leakage, we insert an $LC$ filter in each dc bias line. An $LC$ filter is also used to dc bias the cavity, in contrast to previous devices that used a spiral inductor on the dc tap and no filter along gate bias lines. \cite{Petersson_Nature_2012} Figure~\ref{fig:2}(a) shows the image of a single filter, which consists of a long interdigitated capacitor with $C_\text{f} \approx 1$ pF and a spiral inductor with $L_\text{f} \approx 13$ nH. The overall dimensions of a filter are 700 $\mu$m by 200 $\mu$m. 

To evaluate the attenuation of the filter, we measure its transmission $|S_\text{21}|^2$ at a temperature $T = 1.5$ K. The data, shown in Fig.~\ref{fig:2}(b), display a clear roll-off with frequency $f$. Oscillations with a frequency spacing of $\sim$500 MHz are also seen throughout the data range. The oscillations may be due to reflections at the wire bonds connecting the filter to the circuit board, parasitic modes introduced by discontinuities in the ground plane, and parasitic capacitances/inductances that result from the relatively large size of the filter components. On average, 20 dB of attenuation is obtained around the cavity center frequency $f_\text{c} = 7.67$ GHz [Fig.~\ref{fig:2}(c)]. 

For comparison,  $|S_{21}|^2$, as calculated using the ABCD matrix approach, is plotted in Fig.~\ref{fig:2}(b).\cite{pozar2011microwave} The theory predicts a filter attenuation of 24 dB at $f = 7.67$ GHz. With the exception of the oscillations in the data, the overall transmission through the filter is in good agreement with theory. The undesired oscillations may be suppressed by using air-bridges to better connect distinct regions of the cavity ground plane, and improved circuit board designs to minimize the impedance of the wirebonds. \cite{ChenAPL}

The incorporation of $LC$ filters into the cavity design results in a significant increase in the cavity quality factor (cQED devices of the same DQD design without $LC$ filters have $Q$ $<$ 1,000). Figure \ref{fig:2}(c) shows the normalized cavity transmission $|S_{21}|^2$ as a function of $f$ with the DQD configured in Coulomb blockade at $T = 10$ mK. The input power $P_\text{in} \approx -130$ dBm corresponds to an intra-cavity photon number $n \approx 3$. A fit to a Lorentzian function yields $Q$ = 5,400, corresponding to a total photon loss rate $\kappa / 2 \pi = f_\text{c} / Q = 1.4$ MHz. Using the Sonnet EM simulation program, we estimate the cavity input and output coupling rates to be $\kappa_\text{in} / 2 \pi = \kappa_\text{out} / 2 \pi = 0.4$ MHz. The remaining loss rate of 0.6 MHz may be attributed to a combination of internal loss due to the dielectric layers under the cavity and remnant microwave leakage through the $LC$ filters. The internal loss may be reduced by etching away Al$_2$O$_3$ and Si$_{0.7}$Ge$_{0.3}$ in the gap between the cavity center pin and the ground plane where the electric field intensity is large. Microwave leakage can be further suppressed through improved filter designs such as multi-pole $LC$ filters and band-stop filters. \cite{Houck_BroadbandFilter}

\begin{figure}[t]
	\centering
	\includegraphics[width=\columnwidth]{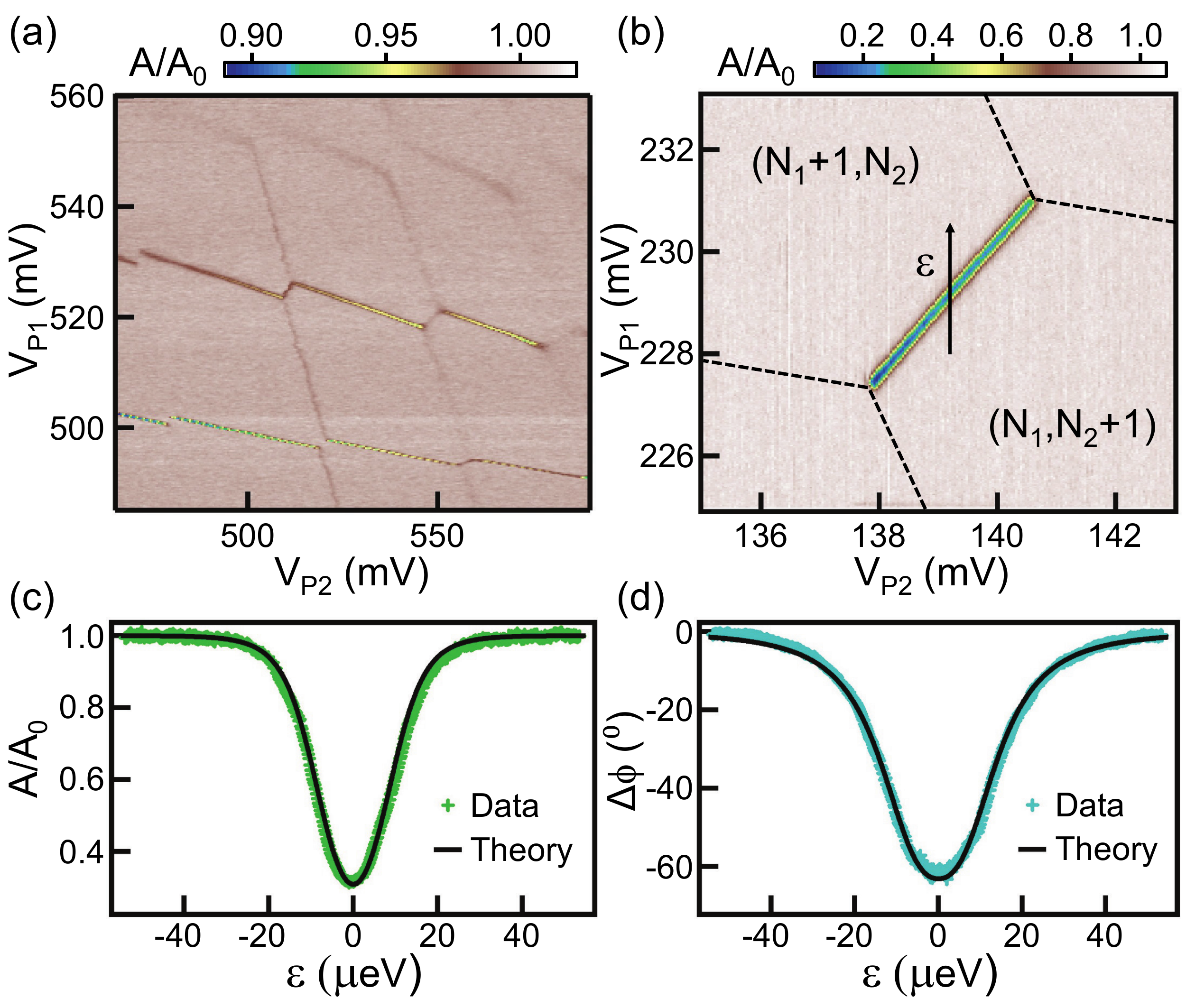}
	\caption{(a) Cavity transmission amplitude, $A/A_0$, measured as a function of the voltages $V_\text{P1}$ and $V_\text{P2}$ on plunger gates P1 and P2. (b) $A/A_0$ at an $(N_1+1,N_2) \leftrightarrow (N_1,N_2+1)$ interdot charge transition. Dashed lines mark the boundaries of the charge stability diagram. (c) $A/A_0$, and (d) the cavity phase response $\Delta \phi$, plotted as a function of $\epsilon$.}
\vspace{0.5cm}
	\label{fig:3}
\end{figure}

We next form a DQD at $T = 10$ mK using the overlapping Al gate architecture [Fig.\ 1(c)] and demonstrate cavity-based charge sensing. The input port of the cavity is driven at a fixed frequency $f = f_\text{c}$ and power $P_\text{in} = -121$ dBm ($n \approx 20$), while the signal exiting the cavity is amplified and demodulated to yield the transmission amplitude $A$ and phase response $\Delta \phi$. \cite{Petersson_Nature_2012} We plot the normalized cavity transmission amplitude $A/A_0$ as a function of $V_\text{P1}$ and $V_\text{P2}$ in Fig.~\ref{fig:3}(a), where the normalization constant $A_0$ is set such that $A/A_0 = 1$ when the DQD is in Coulomb blockade.  The DQD charge stability diagram is revealed in the cavity amplitude response, with the boundaries of charge stability islands delineated by suppressed cavity transmission amplitudes, $A/A_0 < 1$. Here charge transfer between the DQD and the S/D reservoirs results in dispersive shifts of the cavity center frequency, and a reduction in the amplitude of the transmitted signal. \cite{Walraff_2012_PRL,Walraff_PRB_2012_LeadPhysics,Kontos_PRX_2016_Fork}  Dot 1 charge transitions generally have a larger visibility in the data since plunger gate P1 is directly connected to the cavity.

A crucial parameter characterizing cQED systems is the coherent coupling rate between the qubit and the cavity, $g_\text{c}$. A large value of $g_\text{c}$ allows for rapid transfer of quantum states between the qubit and the cavity. Strong coupling is achieved when $g_\text{c}$ exceeds both $\kappa$ and the qubit decoherence rate $\gamma$. \cite{Strong_Coupling_CooperPair,Sillanpaa2007,Majer2007} We estimate $g_\text{c}$ in this device architecture by focusing on an interdot charge transition in the many-electron regime, $(N_1+1,N_2) \leftrightarrow (N_1,N_2+1)$, where $N_1$/$N_2$ denotes the number of electrons in dot 1/dot 2. Charge dynamics in the DQD result in a strong reduction in $A/A_0$ at the interdot charge transition [Fig.~\ref{fig:3}(b)]. At the interdot charge transition, the total number of electrons in the DQD is fixed and a single excess electron functions as a charge qubit with a transition frequency $f_\text{a} = \sqrt{\epsilon^2+4t_\text{c}^2}/h$, where $\epsilon$ is the level detuning, and $t_\text{c}$ is the interdot tunnel coupling. \cite{Petta_Charge_PRL_2004} For this device configuration, the minimum qubit frequency $f_\text{a} = 2t_\text{c}/h$ is close to the cavity frequency $f_\text{c}$, leading to a strong dispersive shift in cavity transmission and the observed reduction in $A/A_0$. At large values of $\epsilon$, $f_\text{a} \gg f_\text{c}$, and the charge qubit becomes decoupled from the cavity.

In Figs.~\ref{fig:3}(c) and 3(d), we plot $A/A_0$ and $\Delta \phi$ as a function of $\epsilon$ for the $(N_1+1,N_2) \leftrightarrow (N_1,N_2+1)$ interdot charge transition. These data are fit to cavity input-output theory \cite{Petersson_Nature_2012,Viennot408} using the measured values of $f_\text{c}$ and $\kappa$. Best fit parameters yield $t_\text{c} = 16.4$ $\mu$eV, $g_\text{c}/2\pi = 23$ MHz, and a charge qubit decoherence rate $\gamma/2\pi = 40$ MHz.  The values of $g_\text{c}$ and $\gamma$ in our hybrid cQED device compare favorably with those in other semiconductor systems. \cite{Walraff_2012_PRL,Toida_GaAs_PRL,Petersson_Nature_2012,Viennot408,Graphene_cQED_2015} The relatively small charge qubit decoherence rate merits further investigation and may be due to the on-chip microwave filters and screening provided by the overlapping Al accumulation gates.  A recent paper by Bruhat \textit{et al.}\ advocates for a reduction in $E_\text{c}$ to minimize sensitivity to charge noise. \cite{KontosSC}  Our results indicate that the highly desired regime of strong-coupling, $g_\text{c} > [\kappa, \gamma]$, could be achieved with only a two-fold reduction in the qubit decoherence rate. 

In conclusion, we have demonstrated readout of a Si/SiGe DQD that is embedded in a superconducting cavity. A quality factor $Q$ = 5,400 is achieved by minimizing photon losses through the use of compact, on-chip $LC$ filters. The DQD stability diagram is visible in measurements of the transmission amplitude of the cavity. Analysis of the cavity response at an interdot charge transition yields a charge-cavity coupling rate $g_\text{c}/2\pi = 23$ MHz. Looking ahead, this hybrid Si/SiGe QD cQED system could be used for the spectroscopy of low-lying valley states in Si, \cite{PhysRevB.94.195305} demonstrations of strong coupling, \cite{Strong_Coupling_CooperPair} and explorations of the resonant exchange regime of triple quantum dots. \cite{Taylor_PRL_2013,Guido_RX_PRB_2015}

\begin{acknowledgements}
We acknowledge valuable discussions with T. M. Hazard, Y.-Y. Liu, and S. Putz and thank J. Kerckhoff for suggesting the $LC$ filters. Research was sponsored by ARO grant W911NF-15-1-0149 and the National Science Foundation (DMR-1409556 and DMR-1420541). The views and conclusions contained in this Letter are those of the authors and should not be interpreted as representing the official policies, either expressly or implied, of the United States Department of Defense or the U.S. Government. Devices were fabricated in the Princeton University Quantum Device Nanofabrication Laboratory.
\end{acknowledgements}

\bibliographystyle{apsrev4-1}
\bibliography{references_v5}

\end{document}